\documentclass[11pt,twoside]{article}
\usepackage{asp2014}

\aspSuppressVolSlug
\resetcounters

\begin{document}

\title{Organics in Disk Midplanes with the ngVLA}
\author{Karin I. \"Oberg$^1$, L. Ilsedore Cleeves$^2$, Ryan Loomis$^3$\\
\affil{$^1$Harvard-Smithsonian Center for Astrophysics, Cambridge, MA 02138; \email{koberg@cfa.harvard.edu}}
\affil{$^2$Department of Astronomy, University of Virginia, 530 McCormick Rd, Charlottesville, VA 22904; \email{lic3f@virginia.edu}}
\affil{$^3$NRAO, 520 Edgemont Rd, Charlottesville, VA 22903; \email{rloomis@nrao.edu}}}

% This section is for ADS Processing.  There must be one line per author.
\paperauthor{Karin I. \"Oberg}{koberg@cfa.harvard.edu}{}{Harvard University}{Department of Astronomy}{Cambridge}{MA}{02138}{U.S.A.}

%Please include a brief abstract that will be used by ADS for searching purposes.  
\begin{abstract}
Planets assemble in the midplanes of protoplanetary disks. The compositions of dust and gas in the disk midplane region determine the compositions of nascent planets, including their chemical hospitality to life. In this context, the distributions of volatile organic material across the planet and comet forming zones is of special interest. These are difficult to access in the disk midplane at IR and even millimeter wavelengths due to dust opacity, which can veil the midplane, low intrinsic molecular abundances due to efficient freeze-out, and, in the case of mid-sized organics, a mismatch between expected excitation temperatures and accessible line upper energy levels. At ngVLA wavelengths, the dust is optically thin, enabling observations into the planet forming disk midplane. ngVLA also has the requisite sensitivity; using TW Hya as a case study, we show that ngVLA will be able to map out the distributions of  diagnostic organics, such as CH$_3$CN, in nearby protoplanetary disks. 
\end{abstract}

\section{Introduction}

A planet's potential hospitality to life or 'habitability' depends on its bulk composition, i.e. whether it is rocky, and surface temperature, i.e. whether it can sustain liquid water. An equally important consideration, however, is whether the planet is chemically hospitable to life or chemically habitable, i.e. whether there is access to water and soluble organic molecules on the planet's surface. Furthermore, some initial organic compositions may be more likely than others to result in a chemistry that eventually brings about an origin of life. Studies of comet compositions reveal that the Solar Nebula from which the Solar System emerged was rich in CH$_3$OH, CH$_3$CHO, CH$_3$CCH, HCOOH, HNCO, HC$_3$N and CH$_3$CN and other intermediate sized organic molecules \citep{Mumma11,LeRoy15}. Several of these, but especially nitriles and unsaturated hydrocarbons, have been found to be particularly well suited to form the building blocks of proteins and RNA \citep{Powner09,Patel15}. Understanding the prevalence of this suite of molecules in analogs to the Solar Nebula is needed to assess how often planets form in a chemical environment that is conducive to producing chemically habitable planets. 

Temperate, habitable zone, planets are expected to form relatively close to their host stars, within a few au around Solar analogs, and obtain their bulk compositions from co-located disk material.  Their volatile budget is set by the direct accretion of this material in the warm inner disk, and by later delivery from asteroids and comets formed at greater distances, up to 10s of au, from the star \citep{Morbidelli12,Raymond14}. Disk chemical compositions out to at least 30 au may thus affect the chemical habitability of nascent planets. It is important to note that we do not expect a disk to present a single composition across the inner 30 au, but rather steep chemical gradients due to radial gradients in temperature,  radiation flux and density \citep{Aikawa99,Willacy00,Oberg11e,Henning13}. With ALMA such chemical gradients are indeed observed (Fig. 1). There is also not a single chemical composition at a set radius due to analogous vertical chemical gradients  \citep{Calvet91,Aikawa99}, i.e. the disk midplanes where planets and planetesimals assemble are often chemically distinct from more elevated disk layers. Most observations to date probe elevated disk layers and the community then uses models to connect these observations to midplane compostions \citep{Oberg15}. 

\section{The role of ngVLA in astrochemical disk explorations}

\articlefigure{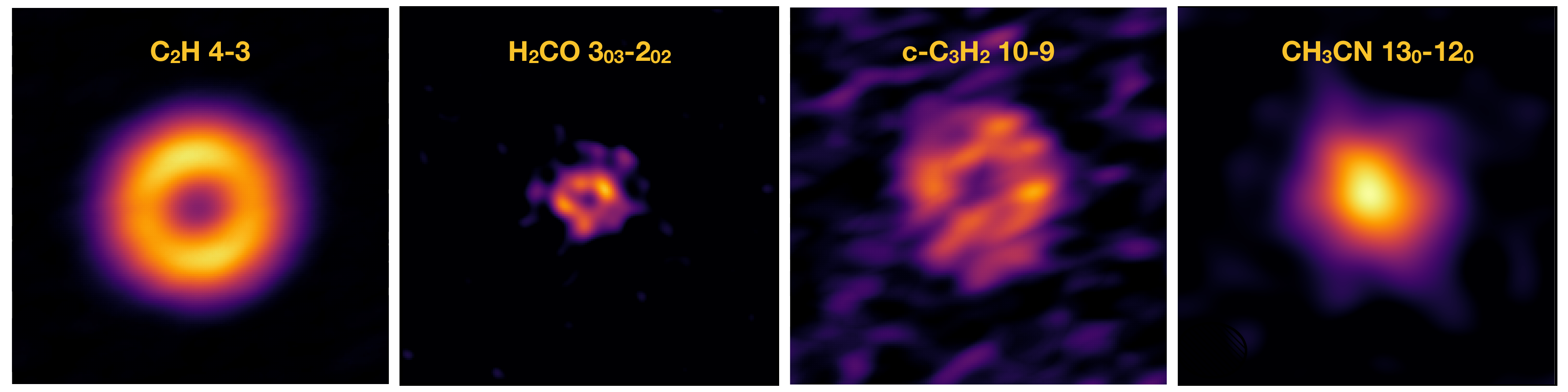}{fig:alma}{ALMA observations of small and mid-sized organics in the disk around nearby T Tauri star TW Hya. C$_2$H and c-C$_3$H$_2$ observations are from \cite{Bergin16}, H$_2$CO from \cite{Oberg17} and CH$_3$CN from \cite{Loomis18b}. Each panel is $\sim5"\times 5"$ and spatial resolutions span $0".4-0".8$}

The chemical composition of Solar Nebula analogs, protoplanetary disks, have been studied at sub-millimeter and millimeter wavelengths for decades, revealing a large number of 2-6 atom organic and inorganic molecules, including the nitriles CN, HCN, HNC, CH$_3$CN, HC$_3$N, hydrocarbons C$_2$H and c-CH$_3$H$_2$, and O-bearing organics H$_2$CO, CH$_3$OH and HCOOH (Fig. 1) and \citep{Dutrey97,Thi04,Oberg11a,Chapillon12,Qi13b,Kastner14,Graninger15,Oberg15,Walsh16,Favre18}. It is becoming increasingly clear, however, that these observations probe the upper layers of disks, and do not reach into the planet forming midplanes \citep{Bergner18,Loomis18b}. First the molecular excitation temperatures in the midplane are expected to be low, and millimeter observations of mid-sized organics such as CH$_3$CN mainly sample the warmer more elevated molecular layers due to relatively high line upper energy levels. Second, and more seriously, high-resolution dust observations with ALMA has revealed that the dust becomes optically thick in the innermost disk \citep{Cleeves16c,Huang18}, and the presence of sub-structure in disks may result in optically thick dust also at larger disk radii \citep{Brogan15,Andrews16}.

At longer wavelengths, dust opacity is less of an issue, and observations of line emission from disk midplanes becomes possible. Compared to millimeter wavelengths, such lines are intrinsically weak, however, and at least an order of magnitude increase in sensitivity is required compared to existing radio facilities, i.e. such observations cannot be carried out with the JVLA, but would be possible with ngVLA for a handful of key molecules in nearby disks. ngVLA would also readily provide the resolution of a fraction of an arc second required to map out the organic content across nearby protoplanetary disks. 

 ngVLA provides access to lines of a range of mid-sized and more complex organics. For the case of disks, whose molecular emission are intrinsically weak, we only consider mid-sized organics. We suggest that the following ladders of three known disk molecules are especially advantageous to target because they present line upper energy levels that match expected disk midplane temperatures -- their upper state energies range between a few and 10s of K -- and cover enough of a temperature range to provide good constraints on the molecular excitation temperatures and column densities: the fundamental 1-0 and 2-1 k-ladder transitions of CH$_3$CN, J=1--0 to 5--4 transitions of HC$_3$N, and $1_{10} - 1_{11}$, $2_{11} - 2_{12}$, $3_{12} - 2_{13}$, $4_{13} - 4_{14}$ transitions of H$_2$CO. The same frequency range,  4--48 GHz, also presents k-ladders of CH$_3$OH, HCOOH and other O-bearing organics, but as explained below it is more speculative whether ngVLA will have sufficient sensitivity to detect them. 

\section{Case study: Mapping organics in the TW Hya protoplanetary disk midplane with ngVLA}

TW Hya hosts the most nearby and well-studied protoplanetary disk \citep{Wilner00,Kastner02,Bergin15,Andrews16,Schwarz16,Huang18}. The disk chemical composition has been constrained at millimeter and sub-millimeter wavelengths \citep{Thi04,Qi08,Qi13c,Kastner14,Cleeves15,Bergin16,Oberg17}, revealing strong emission from CH$_3$CN, and the only detections to date of CH$_3$OH and HCOOH \citep{Walsh16,Favre18,Loomis18b}. In this section we explore what investment of ngVLA time would be needed to characterize the organic chemistry in the TW Hya disk plane and extrapolate to other potential disk targets.

There are two paths to estimating line intensities at ngVLA wavelengths: predictions from disk chemistry models, and extrapolations from line emission observations at shorter wavelengths. We take the former approach, but check it against empirical extrapolations where possible. In particular, we use the disk chemistry model from \cite{Cleeves15}, which was originally optimized to reproduce the emission from small ions in TW Hya, but has also been shown to well reproduce CH$_3$CN emission at millimeter wavelengths \citep{Loomis18b}. 

Figure 2 shows the predicted disk integrated spectra between 1 and 50 GHz for TW Hya with a spectral resolution of 0.2 km/s. The first thing to note is the potential line richness of TW Hya. Many of these lines originate from the mid-size cometary organics that are the molecular targets of this science case. Among the strongest predicted lines are the CH$_3$CN k-ladders around 18 and 37 GHz. Using an empirical extrapolation from existing ALMA observations results in a factor of two higher line intensities, suggesting that the presented model spectra is conservative, at least with respect to CH$_3$CN. Other strong lines can be attributed to CS, C$_2$S, HCOOH, H$_2$CO, NH$_3$, and NH$_2$CN (cyanimide). $>5\sigma$ detections of lines from these species should be possible in a few hours of ngVLA time. We note that while NH$_3$ is not an organic it is an especially interesting target molecule, both because it is a predicted important reservoir of nitrogen and because it could be used to measure the disk gas temperature. More complex molecules than these, as well as CH$_3$OH may be detectable using longer (100~h) integrations and a combination of matched filter and line stacking techniques \citep{Loomis18a}. 

The above disk integrated observations will provide important constraints on disk midplane compositions, but to characterize chemical gradients across the planet and comet forming disk regions require maps at sub-arcsecond resolution. Adopting a spatial resolution of 
0".2, corresponding to 12 au for TW Hya's distance of 59.5 pc \citep{Brown16} we can estimate the sensitivity needed for mapping CH$_3$CN based on existing ALMA observations at 0".8 resolution. At this spatial resolution CH$_3$CN is barely resolved and almost all emission originate from within a disk radius of 0".5. Conservatively we assume that that emission is uniform interior of 0".5 (Fig. 1 shows that sub-structure is likely), and the flux per beam is then 25x less than the disk integrated emission. Based on Fig. 2 the strongest CH$_3$CN lines will have a peak flux of 10 mJy / 25 $\sim 400 \mu$Jy per beam, and a 5$\sigma$ ngVLA map should be achievable in $\sim$10 hours of integration. 

TW Hya is unique in its proximity and therefore in its line strengths. To extend observations to a sample of disks we would need to observe disks in nearby star forming regions, such as Taurus, at distances of $\sim$150 pc, or 2.5x farther away than TW Hya. Based on millimeter observations, line emission from these disks are on the order of 2--5x weaker compared to TW Hya, requiring on average a factor of 10 longer integrations. To map the brightest organics in these disks would thus require $\sim$100~h per disk. 

\articlefigure{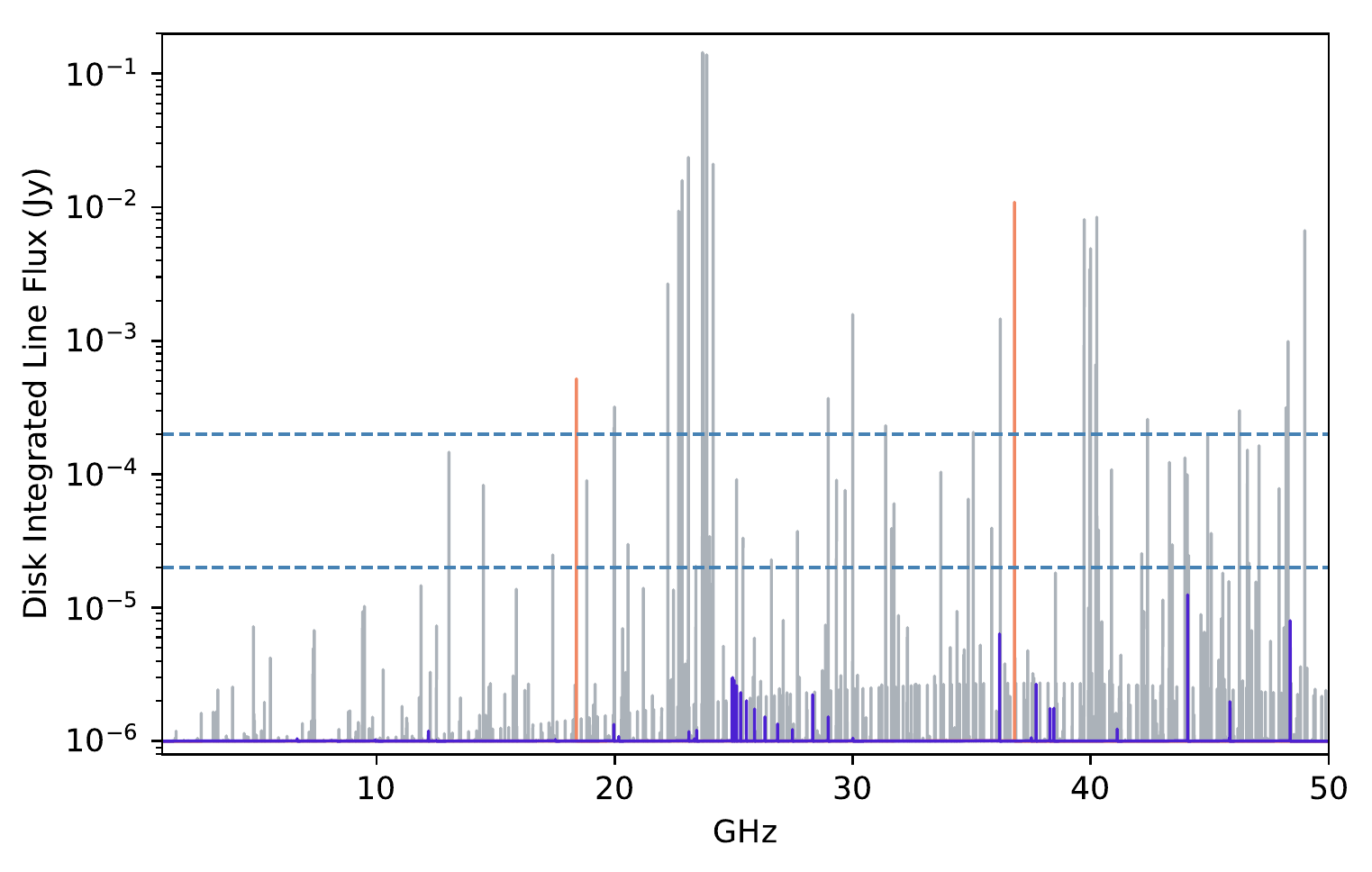}{fig:spec}{Predicted disk integrated spectra towards TW Hya between 1 and 50 GHz. The model spectra is generated using the disk chemistry model for TW Hya presented in \cite{Cleeves15}. The two dashed lines show the sensitivity limits for ngVLA after 1 and 100 h integration, using the baseline sensitivity of ngVLA 30 microJy in 10 km/s in one hour and a required spectral resolution of 0.2 km/s. CH$_3$CN is marked in orange and CH$_3$OH in blue.}

\section{Synergies and concluding remarks}

Observations of organic molecule abundances and distributions in protoplanetary disks are key to calibrate astrochemical disk models, and to predict how often we can expect planets to form in environments that are chemically conducive for the emergence of life. ngVLA will provide a crucial part of the puzzle by giving access to the disk midplanes where planets and comets assemble. ALMA is currently providing the complementary information on molecular distributions at scales of 10-100 a.u. in more elevated disk layers, and JWST will soon give access to the upper layers of the innermost disk regions. Together these three facilities will provide orthogonal constraints on the chemistry of planet forming disks and aid astrochemists in developing a complete picture of disk chemistry and the chemistry of nascent planets.

%\acknowledgements ...  % Keep this text on the same line as the \verb"\acknowledgements" command because it makes things a lot easier.

%\bibliography{../../BIB/mybib}  % For BibTex

% For non-BibTex:

\end{document}